\documentclass[12pt,nohyper,notoc]{JHEP}

\usepackage{amsmath,array,amssymb,cite} 

\def\elabel#1{\label{#1}}

\DeclareSymbolFont{AMSb}{U}{msb}{m}{n}
\DeclareSymbolFontAlphabet{\Bbb}{AMSb}

\def\rsen{\setcounter{equation}{0}}

\def\Z{{\cal Z}}
\def\N{{\cal N}}
\def\Nf{{N_{\rm F}}}
\def\det{{\rm det}}

\def\tr{{\rm tr}}

\def\Qtil{\tilde{Q}}

\def\Adagtil{{\tilde{A}}^{\dagger}}

\def\Weff{{\cal W}_{\rm eff}}
\def\Wads{{\cal W}_{\rm \scriptscriptstyle ADS}}

\def\L{{\cal L}}
\def\Leff{{\cal L}_{\rm eff}}

\def\thetabar{\bar{\theta}}

\def\sqrtwo{\sqrt{2}\,}
\def\hf{{\textstyle{1\over2}}}

\def\sigmabar{\bar\sigma}

\def\Tr{{\rm Tr}}

\def\N{{\cal N}}

\def\lambdabar{\bar\lambda}

\def\sst{\scriptscriptstyle}
\def\st{\scriptscriptstyle}

\def\D{{\cal D}}
\def\Dbarslash{\,\,{\raise.15ex\hbox{/}\mkern-12mu {\bar\D}}}
\def\delslash{\,\,{\raise.15ex\hbox{/}\mkern-9mu \partial}}
\def\Dslash{\,\,{\raise.15ex\hbox{/}\mkern-12mu \D}}

\def\sigmabar{\bar\sigma}
\def\psibar{\bar\psi}

\def\N{{\cal N}}

\def\hf{{\textstyle{1\over2}}}

\def\sqrtwo{\sqrt{2}\,}

\def\hf{{\textstyle{1\over2}}}

\def\det{{\rm det}}

\def\sqrtwo{\sqrt{2}\,}

\def\N{{\cal N}}

\def\sigmabar{\bar\sigma}

\def\VEV#1{\left\langle #1\right\rangle}
\def\Vev#1{\big\langle{#1}\big\rangle}

\def\sigmabar{\bar\sigma}

\def\N{{\cal N}}

\def\hf{{\textstyle{1\over2}}}

\def\lambdabar{\bar\lambda}
\def\psibar{\bar\psi}
\def\sqrtwo{\sqrt{2}\,}

\def\uA{\,\lower 1.2ex\hbox{$\sim$}\mkern-13.5mu A}
\def\uphi{\,\lower 1.2ex\hbox{$\sim$}\mkern-13.5mu \phi}
\def\uX{\,\lower 1.2ex\hbox{$\sim$}\mkern-13.5mu X}
\def\uD{\,\lower 1.2ex\hbox{$\sim$}\mkern-13.5mu {\rm D}}

\def\uF{\,\lower 1.2ex\hbox{$\sim$}\mkern-13.5mu F}
\def\uW{\,\lower 1.2ex\hbox{$\sim$}\mkern-13.5mu W}
\def\uWbar{\,\lower 1.2ex\hbox{$\sim$}\mkern-13.5mu {\overline W}}

\def\uV{\,\lower 1.2ex\hbox{$\sim$}\mkern-13.5mu V}
\def\uv{\,\lower 1.0ex\hbox{$\scriptstyle\sim$}\mkern-11.0mu v}
\def\uPsi{\,\lower 1.2ex\hbox{$\sim$}\mkern-13.5mu \Psi}
\def\uPhi{\,\lower 1.2ex\hbox{$\sim$}\mkern-13.5mu \Phi}
\def\uchi{\,\lower 1.5ex\hbox{$\sim$}\mkern-13.5mu \chi}
\def\Psibar{\bar\Psi}
\def\uPsibar{\,\lower 1.2ex\hbox{$\sim$}\mkern-13.5mu \Psibar}
\def\upsi{\,\lower 1.5ex\hbox{$\sim$}\mkern-13.5mu \psi}
\def\psibar{\bar\psi}
\def\upsibar{\,\lower 1.5ex\hbox{$\sim$}\mkern-13.5mu \psibar}
\def\upsibarzero{\,\lower 1.5ex\hbox{$\sim$}\mkern-13.5mu \psibar^\zero}
\def\ulambda{\,\lower 1.2ex\hbox{$\sim$}\mkern-13.5mu \lambda}
\def\ulambdabar{\,\lower 1.2ex\hbox{$\sim$}\mkern-13.5mu \lambdabar}
\def\ulambdabarzero{\,\lower 1.2ex\hbox{$\sim$}\mkern-13.5mu \lambdabar^\zero}
\def\ulambdabarnew{\,\lower 1.2ex\hbox{$\sim$}\mkern-13.5mu \lambdabar^\new}
\def\D{{\cal D}}

\def\N{{\cal N}}
\def\Dslash{\,\,{\raise.15ex\hbox{/}\mkern-12mu \D}}
\def\Dbarslash{\,\,{\raise.15ex\hbox{/}\mkern-12mu {\bar\D}}}
\def\delslash{\,\,{\raise.15ex\hbox{/}\mkern-9mu \partial}}
\def\delbarslash{\,\,{\raise.15ex\hbox{/}\mkern-9mu {\bar\partial}}}

\title{On Affleck-Dine-Seiberg Superpotential
and Magnetic Monopoles in Supersymmetric QCD}

\author{N. Michael Davies and Valentin V. Khoze\\
Department of Physics, University of Durham,
Durham, DH1 3LE, UK\\
E-mail:{\tt n.m.davies@durham.ac.uk}, {\tt valya.khoze@durham.ac.uk}}

\abstract{Certain exact results in supersymmetric gauge theories are
generated by non-perturbative effects different from instantons.
In supersymmetric QCD with $N$ colours and
$\Nf$ fundamental flavours we examine the Affleck-Dine-Seiberg 
(ADS) superpotential using controlled semi-classical analysis.
We show how for $\Nf<N-1$ the ADS superpotential 
arises from monopole contributions to the path integral of the supersymmetric
gauge theory compactified on ${\Bbb R}^3 \times S^1$.
These are the monopole effects leading to gaugino condensation and confinement
of the low-energy $SU(N-\Nf)$ supersymmetric gauge theory.}

\keywords{Supersymmetry, Monopoles and Instantons, Superpotentials, 
Compactification}

\preprint{{\tt hep-th/9911112}}

\begin{document}

\section{Introduction and Motivation}

One of the famous exact results in supersymmetry
is the Affleck-Dine-Seiberg (ADS) superpotential \cite{ADS} generated
in supersymmetric QCD with $\Nf$ fundamental
flavours and gauge group $SU(N)$
\begin{equation}
\Wads^{\Nf,N} \ = \
(N-\Nf)\,\left({\Lambda^{3N-\Nf}_{\Nf,N}\over
\det_{\Nf}\big(\Qtil^{r} Q_s\big)}\right)^{1\over N-\Nf} 
\ .
\elabel{finfin}\end{equation}
It is well known that for $\Nf<N-1$ the ADS superpotential receives
no contributions from instantons, and 
arises from non-perturbative effects associated with gaugino condensation 
in the unbroken $SU(N-\Nf)$ gauge group \cite{ADS}.
In our earlier work \cite{mono1} we showed that 
gaugino condensation is generated
by monopole contributions to the path integral of supersymmetric
gauge theories compactified on ${\Bbb R}^3 \times S^1$.
In this paper we make an obvious link between these two ideas \cite{ADS,mono1} 
and explicitly derive the ADS superpotential 
in the background of $N-\Nf$ varieties of $SU(N-\Nf)$ monopoles embedded
in the $SU(N)$ supersymmetric QCD with $\Nf$ flavours. We will also
explain why this semi-classical derivation is valid in the a priori
strongly coupled theory. For related work on ${\Bbb R}^3 \times S^1$
in a different context see \cite{SWthree, AHISS, Dorey}.

The last few years have seen an impressive agreement between 
known exact results in supersymmetric gauge theories and direct
field-theoretical calculations. In most of the cases considered, the
non-perturbative information contained in the exact results 
\cite{SW,AdS,BG}
was
restricted to multi-instanton contributions \cite{MO,KMS,DHKMV}. 
At the same time, some of the known exact results 
in supersymmetric gauge theories are
generated by non-perturbative effects different from instantons.
We expect that these non-perturbative effects
can still be evaluated and understood semi-classically, but
in slightly unusual settings. The motivation of this letter and of
its predecessor Ref.~\cite{mono1} is to show how this works for two important
examples: gaugino condensation and the ADS superpotential. 
In both cases there is a non-Abelian, asymptotically free (sub)group 
which makes the theory strongly coupled. This is an obstacle for a direct application of semi-classical analysis, which makes these cases
qualitatively different from the weakly coupled scenarios:
\newline
{\it 1.} In the Seiberg-Witten case \cite{SW} the
$SU(N)$ gauge group is broken by the Higgs mechanism
to the Abelian subgroup $U(1)^{N-1}$ with effective couplings determined
by the VEVs. The holomorphic nature of this dependence, guaranteed
by non-renormalization theorems in $\N=2$, then allows one to travel
smoothly to an
arbitrarily weak coupling regime saturated by the relevant
semi-classical physics: perturbation theory, and multi-instanton effects 
calculated in \cite{MO,KMS}.
\newline
{\it 2.} In general, when there is no unbroken non-Abelian subgroup,
the F-terms in any $\N=1$ supersymmetric theory can be usually determined
with a constrained instanton calculation
\cite{ADS,NSVZtwo,Fuchs:1986ft,Cordes,FP} as reviewed
in \cite{Shifman:1999mv}. A simple example of this set-up is the
supersymmetric QCD with $\Nf=N-1$.
\newline
{\it 3.} In the AdS/CFT programme \cite{AdS} the non-Abelian
gauge group is unbroken, but the $\beta$-function is zero,
and for successful multi-instanton calculations
\cite{BG,DHKMV,HKM1,spnt,HKM2},
the constant coupling can be taken as small as needed.

It has been suspected for a long time \cite{BFST} 
that in the strongly coupled theories, 
instantons should be thought of as composite states of more basic
configurations, the `instanton partons'. These partons
would give important contributions to 
the non-perturbative dynamics at strong coupling. 
In Ref.~\cite{mono1} we identified the instanton
partons with the $N$ varieties of monopoles which appear when Minkowski
space-time of $SU(N)$ supersymmetric gauge theory
is partially compatified on ${\Bbb R}^3 \times S^1$. This compactification
is required for two reasons. First, because of the finite radius $\beta$ 
of $S^1$,
the monopole solutions have finite action, 
which makes them important semi-classically.
Second, the gauge fields with Lorentz index referring to the $S^1$ 
direction develop
vacuum expectation values \cite{mono1}
\begin{equation}
\langle A_4 \rangle \ = \ 
 {\rm diag}\Big({N-1 \over  N}{\pi \over i \beta} \ , 
\  {N-3 \over  N}{\pi \over i \beta} \ ,
\ldots \ , \ 
 -{N-1 \over  N}{\pi \over i \beta}
\Big) \ ,
\elabel{lgenn}\end{equation}
which break the non-Abelian $SU(N)$ to the Abelian subgroup 
$U(1)^{N-1}$. 
As the VEVs are inversely proportional to the radius $\beta$,
the theory becomes weakly coupled at small $\beta$ and can be analysed
semi-classically. To return to the strongly coupled theory in
Minkowski space, we need to consider the opposite limit of large $\beta$.
Since all the F-terms are holomorphic functions of the fields and since
the VEVs of the fields \eqref{lgenn} are holomorphic functions of $\beta$,
the power of holomorphy \cite{Seiberg93} allows to analytically continue
the semi-classical values of the F-terms to the strong-coupling regime. 

In supersymmetric QCD with $N$ colours and $\Nf$ fundamental flavours in the
Higgs phase, the gauge group is broken by the matter VEVs to $SU(N-\Nf)$.
In Section II we will review the 
monopole calculus in the theory with $\Nf=0$
on ${\Bbb R}^3 \times S^1$. The monopole effects in this theory
generate the gaugino condensate and give a mass to the dual (magnetic)
photon which implies confinement of electric charges.
In Section III we will include the
matter fields and start in uncompactified Minkowski space.
We will first integrate out the classically massive (Higgsed)
fields. The resulting effective lagrangian will contain the massless
mesons $M=\Qtil Q$ and the strongly-interacting 
massless $SU(N-\Nf)$ gauge
supermultiplet, $W_{\st SU(N-\Nf)}$. 
These two sectors will be coupled to each other
in the effective lagrangian via non-renormalizable interactions generated
perturbatively by integrating out massive fields propagating in the loops.
The next step
will be to integrate out the gauge supermultiplet, $W_{\st SU(N-\Nf)}$. 
This is achieved in the monopole background in the $SU(N-\Nf)$ theory
compactified on ${\Bbb R}^3 \times S^1$. 
The superpotential for mesons -- the ADS superpotential 
\eqref{finfin} -- arises from the
gauge-meson interactions when the gauge supermultiplet is integrated out.
Hence, the ADS superpotential \eqref{finfin} is a combination of 
perturbative and monopole effects.
The analysis of Section III can be schematically represented as
\begin{equation}\begin{split}
&\int\D W_{\st SU(N)}\,\D Q\,\D \Qtil\, \ 
e^{i\int d^4x\,\L(W_{\st SU(N)},Q,\Qtil)} \\
 &= \ \int\D W_{\st SU(N-\Nf)}\,\D M\, \ 
e^{i\int d^4x\,\Leff^{(1)}(W_{\st SU(N-\Nf)},M)} \ = \ \int\D M\,
 \ e^{i\int d^4x\,\Leff^{(2)}(M)}
\ .
\elabel{schem}\end{split}\end{equation}
We stress that Eq.~\eqref{schem} represents a 
{\it direct} evaluation of the ADS superpotential contained in
$\Leff^{(2)}(M)$. In particular we are not apealing to renormalization group
arguments relating expressions for different values of $\Nf$.
None of the meson degrees of freedom were integrated out on the
right hand side of Eq.~\eqref{schem}, and the F-term of
$\Leff^{(2)}(M)$ is the definition of the general ADS superpotential
$\Wads^{\Nf, N}$.

\rsen

\section{Monopoles in SYM on ${\Bbb R}^3 \times S^1$}

In this section we consider the pure
$\N=1$ supersymmetric Yang-Mills theory coupled to a background
chiral superfield $T^{(\mu)}$
\begin{equation}
\Z(T^{(\mu)}) \ = \ \int\D W \,
e^{i\int d^4x\,\L(W ,T^{(\mu)})}
 \ = \ 
e^{ -i\left(\int d^4x\,d^2\theta \ \Weff(T^{(\mu)}) + {\rm h.c.}\right)}
\ .
\elabel{sone}\end{equation}
Here $W$ denotes the $SU(N)$ field-strength chiral superfield, and
$T^{(\mu)}$ is a non-dynamical background chiral superfield
\begin{equation}
T^{(\mu)}(y,\theta)\ \equiv\
\tau(y)+\sqrtwo\theta^\alpha\chi^\tau_\alpha(y) +\theta^2F^\tau(y)\ ,
\qquad
\langle \tau \rangle \ = \ {4\pi i\over g^2 (\mu)}+{\theta(\mu)\over2\pi}
\elabel{stw}\end{equation}
The superscript $\mu$ indicates the (high) energy scale where $T^{(\mu)}$
becomes dynamical.
The microscopic lagrangian defining the theory reads
\begin{equation}
\L \ = \ {1 \over 4 \pi} \ {\rm Im} \ 
\tr_N
\int d^2\theta \ T^{(\mu)} \ W^{\alpha} W_\alpha 
\ .
\elabel{sfou}\end{equation}
The dynamical quantities we are after are the effective superpotential
$\Weff(T^{(\mu)})$ defined in Eq.~\eqref{sone}
and the gaugino condensate $\langle \tr_N \lambda^2 \rangle$ in the
presence of the $\tau$ field. The two quantities are related due
to a functional identity
\def\condition{\,{\Big|}_{T^{(\mu)}(y,\theta)=\tau}}
\begin{equation}
\VEV{\tr_N\lambda^2}\ =\ {8\pi\over\Z(T^{(\mu)})}{\delta\over\delta
F^\tau(x)}\,\Z(T^{(\mu)})\condition \ = \ 
-8 \pi i \ {\partial \Weff(\tau) \over \partial \tau}
\ ,
\elabel{Zone}\end{equation}
which trivially follows from Eqs.~\eqref{sone}-\eqref{sfou}.

\newpage
{\bf Gauge Theory on ${\Bbb R}^3 \times S^1$}

Let
$x_4$ be periodic with period $\beta/2 \pi$.\footnote{The indices run
over $m=1,2,3,4$ and $\mu=1,2,3$. Our 
conventions are the same as in Ref.~\cite{mono1}.} We must then impose 
periodic boundary conditions for bosons, 
$A_m (x_\mu,x_4=0)=A_m (x_\mu,x_4=\beta)$ and fermions,  
$\lambda (x_\mu,x_4=0)=\lambda (x_\mu,x_4=\beta)$
to preserve supersymmetry. In addition, the local gauge group itself 
must also be composed of gauge transformations periodic on $S^1$:
$U(x_\mu,x_4=0)=U(x_\mu,x_4=\beta)$.
We now summarize the results of the analysis carried out in Ref.~\cite{mono1}.

{\bf 1.} There are $N$ classically flat directions of the vacuum
moduli space of the compactified theory parametrized by
$\langle A_4 \rangle = {\rm diag}(a_1,a_2,\ldots,a_N)$.
This classical moduli space is lifted non-trivially at the quantum level to
\begin{equation}
\langle A_4 \rangle \ = \ 
 {\rm diag}\Big({N-1 \over  N}{\pi \over i \beta} \ , 
\  {N-3 \over  N}{\pi \over i \beta} \ ,
\ldots \ , \ 
 -{N-1 \over  N}{\pi \over i \beta}
\Big) \ ,
\elabel{lgen}\end{equation}
by the monopole-generated superpotential derived in the section III
of ~\cite{mono1}. 
The distinctive feature of \eqref{lgen} is the constant equal spacing
between the VEVs $a_j$.

{\bf 2.} The semi-classical physics of the 
${\Bbb R}^3 \times S^1$ $SU(N)$ theory is described by configurations of
monopoles of $N$ different types.
In general, 
the field configurations of the 
${\Bbb R}^3 \times S^1$ theory which are relevant in the semi-classical regime
are both instantons and monopoles. Remarkably, the instantons on 
${\Bbb R}^3 \times S^1$ can be understood as composite configurations
of $N$ single monopoles, one of each of the $N$ different types
\cite{Nahmtwo,LLY,KvB}. One expects in
an $SU(N)$ theory on ${\Bbb R}^4$ that the lowest charged monopoles come in
$N-1$ different varieties, carrying a unit of magnetic charge from
each of the $U(1)$ factors of the $U(1)^{N-1}$ gauge group left
unbroken by the VEVs. 
The additional monopole, needed to make up the $N$  types, 
is specific to the compactification on ${\Bbb R}^3 \times S^1$
\cite{LLY,KvB}.
The magnetic charge of the new monopole is such that
when all $N$ types of monopoles are present, the magnetic charges
cancel and the resulting configuration only carries a unit instanton
charge. 

{\bf Monopole calculus in SU(2)}

The standard BPS monopole solution in Hedgehog gauge \cite{BogPS} is
\begin{equation}
A^{\sst\rm BPS}_4 (x_\nu)\ = \ \big(v|x| \ {\rm coth}(v|x|) -1
\big) 
{x_a \over |x|^2}{\tau^a \over 2i} \ , \quad
A^{\sst\rm BPS}_\mu (x_\nu)
 \ = \ \Big(1-{v|x| \over {\rm sinh}(v|x|)}
\Big)\epsilon_{\mu\nu a}{x_\nu \over |x|^2}{\tau^a \over 2i}
 \ .
\elabel{bpscn}\end{equation}
These expressions are obviously independent of the 
$S^1$ variable $x_4$,
since the latter can be thought of as the time coordinate of 
the static monopole.
The boundary values of \eqref{bpscn} as $|x| \to \infty$, when gauge
rotated to unitary (singular) gauge, agree with 
\eqref{lgen} for $v=\pi /\beta$
\begin{equation}A^{\sst\rm BPS}_4 \ \to \ v{\tau^3 \over 2i} \ = \ 
{\pi \over \beta} {\tau^3 \over 2i}
 \ = \ \langle A_4 \rangle \ . \elabel{bcsi}\end{equation}
The monopole solution \eqref{bpscn} satisfies the self-duality equations
and has topological charge
\begin{equation}Q \ \equiv \ {1 \over 16 \pi^2} \ \int_0^\beta dx_4 \int d^3 x
\ \tr\,{}^*F_{mn}F^{mn} \ = \ {\beta v \over 2 \pi} \ = \ {1 \over 2}
 \ . \elabel{topq}\end{equation}
The monopole has magnetic charge one,
instanton charge zero, and the action $S_{\sst\rm BPS}$ is
\begin{equation}S_{\sst\rm BPS} \ = \ {4\pi \over g^2(\mu)} \beta v \ = \ 
{4\pi^2 \over g^2(\mu)} \ , \qquad
S_{\sst\rm BPS}(T^{(\mu)}) \ = \ 
-i \pi T^{(\mu)}
\ .
\elabel{macnone}\end{equation}
The solution \eqref{bpscn} precisely two adjoint fermion zero modes 
$\lambda^{\sst\rm BPS}_\alpha = \hf \xi_\beta
(\sigma^m \sigmabar^n)_\alpha^{\ \beta} F^{\sst\rm BPS}_{mn}$,
with normalization \cite{DKMTV}:
\begin{equation}\int d^3 X \int d^2 \xi  \ 
{\rm tr}\left( \lambda^{{\sst\rm BPS} \ \alpha}(x) 
\lambda^{\sst\rm BPS}_\alpha(x) \right) \ = \ {2g^2 S_{\sst\rm BPS}\over\beta}
 \ = \ {8\pi^2  \over \beta}
\ .
\elabel{nfzm}\end{equation}
Here $\sigma^m$ and $\sigmabar^n$ are the four Pauli matrices and 
$\xi_\beta$ is the two-component Grassmann collective coordinate.
The semi-classical integration measure
of the standard single-monopole on ${\Bbb R}^3\times S^1$ reads \cite{DKMTV}:
\begin{equation}
\int d\mu^{\sst\rm BPS} \ = \ \mu^3 \ 
e^{-S_{\sst\rm BPS}(T^{(\mu)})} \
\int {d^3 X \over (2\pi)^{3/2}} \ 8\pi^3 \ 
\int_0^{2\pi} {d \Omega \over \sqrt{2\pi}}{2 \pi \over v} \
\int d^2 \xi {1 \over 8\pi^2} \ .\elabel{msst}\end{equation}
This measure is obtained in the standard way by changing variables
in the path integral from field-fluctuations around the monopole
to the monopole's collective coordinates.
The UV-regularized
determinants over non-zero eigenvalues of the
quadratic fluctuation operators cancel between fermions and bosons
due to supersymmetry.
The ultra-violet divergences are regularized in the Pauli-Villars scheme,
which explains the appearance of the Pauli-Villars mass scale
$\mu$.
We can now compute the single monopole contribution 
to $\Vev{\tr \lambda^2}$ as in Ref.\cite{mono1}:
\begin{equation}
\VEV{\tr \lambda^2}_{\sst\rm BPS} 
 \ =\ \int d\mu^{\sst\rm BPS} \ {\rm tr}\left( 
\lambda^{{\sst\rm BPS} \ \alpha}(x) 
\lambda^{\sst\rm BPS}_\alpha(x) \right) \ = \ 
8 \pi^2  \mu^3 \exp[i\pi T^{(\mu)}]
\ .
\elabel{glbp}\end{equation}

As explained in Ref.~\cite{mono1} the monopole of the second type
will give a contribution identical to \eqref{glbp}.
Putting the two together we get
\begin{equation}
\VEV{\tr \lambda^2} \ = \  16\pi^2 \ \mu^3 \exp[i\pi T^{(\mu)}]
 \ . \elabel{lcstwo}\end{equation}
Substituting \eqref{lcstwo} into \eqref{Zone} gives the equation
for the superpotential $\Weff(\tau)$ with the solution
\begin{equation}
\Weff(T^{(\mu)}) \ = \ 2\mu^3 \ \exp[\pi i T^{(\mu)}]
 \ . \elabel{sfin}\end{equation}
The expressions 
\eqref{lcstwo} and
\eqref{sfin} are, in spite of appearences, renormalization group invariant
and $\mu$-independet. We also note that the dependence on the $S^1$
radius $\beta$ disappeared, and the decompactification limit 
$\beta\to \infty$ does not change the final results \eqref{lcstwo} and
\eqref{sfin}.

{\bf Generalization to SU(N)}

The calculation of the gaugino condensate and the superpotential can be
straightforwardly generalized to the 
case of gauge group $SU(N)$. The quantum vacuum has
\begin{equation} a_{j} - a_{j+1} \ = \ {2\pi \over iN\beta} \ {\rm
mod} \ {2\pi\over i\beta}\ ,
\qquad j=1,2,\ldots,N \ , 
\elabel{lnnn}\end{equation}
and each of the $N$ types of monopoles has equal actions and equal topological charges:
\begin{equation}
S_{\rm mono} \ = \ {8 \pi^2 \over N g^2(\mu)} \ , \qquad
S_{\rm mono} (T^{(\mu)}) \ = \ -{2\pi i \over N} \ T^{(\mu)}
 \ , \qquad
Q_{\rm mono} \ = \ {1\over N }
 \ .\elabel{smon}\end{equation}
The contribution of $N$ monopoles to the gaugino condensate 
and the superpotential is straightforward:
\begin{equation}
\VEV{\tr \lambda^2} \ =
16\pi^2\ 
\mu^3 \ \exp\left[{2\pi i \over N} T^{(\mu)}\right] \ , \quad
\Weff(T^{(\mu)}) \ = \ N \ \mu^3 \ \exp\left[{2\pi i \over N} T^{(\mu)}\right]
 \ . \elabel{sssfin}\end{equation}
This concludes our analysis of supersymmetric QCD with
$\Nf=0$.

\rsen

\section{ADS Superpotential in Supersymmetric QCD}

The microscopic theory is defined in terms
of the $SU(N)$ vector superfield coupled to $\Nf$ chiral superfields
in the $[N]$ reperesentation, $Q_{ur}$ $(u=1,\ldots, N; \ r=1,\ldots, \Nf)$,
and to $\Nf$ chiral superfields
in the $[\bar{N}]$ reperesentation, $\Qtil^{ru}$.
The global classical symmetry of the lagrangian is 
\begin{equation}
SU(\Nf)_{\rm left} \times SU(\Nf)_{\rm right} \times U(1)_{\rm V}\times 
U(1)_{\rm A}\times U(1)_{\rm R} \ , 
\elabel{gcsl}\end{equation}
where $U(1)_{\rm R}$ is the anomaly-free combination of the R-symmetry
and the axial $U(1)_{\rm A}$:
\begin{equation}
W(\theta) \to e^{-i\alpha} W(e^{i\alpha}\theta) \ , \quad
 Q(\theta) \to e^{i\alpha (N-\Nf)/\Nf} Q(e^{i\alpha}\theta) \ ,
\quad \Qtil(\theta) \to e^{i\alpha (N-\Nf)/\Nf} \Qtil(e^{i\alpha}\theta) \ .
\elabel{rsym}\end{equation}
At the quantum level
the classical $U(1)_{\rm A}$ symmetry is anomalous,
and the global {\it quantum} symmetry of the lagrangian is
$SU(\Nf)_{\rm left} \times SU(\Nf)_{\rm right} \times U(1)_{\rm V}\times 
U(1)_{\rm R}$.

The classical vacuum state is determined in the 
standard way via the D-flatness condition and can be brought to a simple `rectangular diagonal' form:
\begin{equation}
\langle A_{ur} \rangle \ = \ \Big\{{\delta_{ur} v_r \ , \ 
{\sst u=1,\ldots,\Nf }
\atop 0  \ , \ {\sst u=\Nf+1,\ldots,N}} \ = \ \langle \Adagtil_{ur} \rangle \ .
\elabel{clv}\end{equation}
The $\Nf$ complex vacuum expectation values $v_1,\ldots, v_\Nf$ 
are not fixed by the
classical lagrangian, and parameterize the $\Nf$-complex-dimensional
classical vacuum moduli space of the theory.
It is now straightforward to determine which symmetries of
the lagrangian are left unbroken by the classical vacuum \eqref{clv}.
The $SU(N)$ gauge symmetry is spontaneously broken by \eqref{clv}
to $SU(N-\Nf)$. In order to determine the surviving global symmetry
it is convenient to restrict ourselves to the case of all VEVs equal,
$v_1=v_2=\ldots \equiv v$.
This choice will not affect the counting of classically
massless degrees of freedom, but will simplify the symmetry reasoning.
This vacuum state breaks the $U(1)_{\rm A}$
and $U(1)_{\rm R}$. It also breaks 
$SU(\Nf)_{\rm left} \times SU(\Nf)_{\rm right}$ to the diagonal 
$SU(\Nf)$ subgroup that rotates $A_{r}$ and $\Adagtil_{r}$ in the
same way and is further
compensated by the gauge transformation, such that the vacuum
 is unchanged.
Thus, the global symmetry respected by the vacuum is 
$SU(\Nf) \times U(1)_{\rm V}$.
Hence, we expect $\Nf^2+1$ massless Goldstone bosons
at the perturbative level,
i.e. $\Nf^2+1$ real massless scalars coming from the broken generators
of the global  $SU(\Nf)\times U(1)_{\rm A} \times U(1)_{\rm R}$.
When the non-perturbative effects are taken into account one of these
massless degrees of freedom will acquire a mass due to the
$U(1)_{\rm A}$ anomaly, and  $\Nf^2$ real scalar degrees of freedom will
remain massless.

At the same time,
the number of {\it classically massless} real scalar degrees of freedom
is $2\Nf^2$. This is twice the number of exact
Goldstone bosons.
The remaining $\Nf^2$ classically massless real scalars must be lifted
by a non-perturbatively generated superpotential.
The functional form of this superpotential Eq.~\eqref{finfin}
was uniquely determined in \cite{ADS}.

The ADS superpotential appears in the low-energy effective description 
of the microscopic theory with $\Nf\le N-1$.
It is important to distinguish between the two cases: $\Nf = N-1$
where the gauge group is completely broken by the vacuum; and  
$\Nf < N-1$ where the non-Abelian gauge subgroup $SU(N-\Nf)$ is
unbroken. The first case is relatively simple and is well understood
\cite{FP,Cordes,ADS}.
For $\Nf = N-1$ the superpotential is generated at the 1-instanton level,
and since the gauge group is completely broken, the instanton calculation
is reliable and infra-red safe. 
In the second, more general case, $\Nf < N-1$, instantons are known to
give trivially vanishing conributions to \eqref{finfin}, nevertheless
the renormalization group decoupling argument from $\Nf = N-1$ to
$\Nf < N-1$ requires the superpotential to be non-vanishing, and
determines the normalzation consant in \eqref{finfin}
 \cite{Seiberg93,FP,Cordes} to be 
$N-\Nf$.

{\bf $\Nf < N-1$: Step One}

For $\Nf < N-1$ there is an unbroken gauge subgroup $SU(N-\Nf)$
and as discussed
in the Introduction, it is important
to realise that the generation of the ADS superpotential is the two-step
process represented by Eq.~\eqref{schem}.
We first look at the perturbative decoupling of the massive vector
bosons and Higgs bosons.
This step is the generalization of the Affleck-Dine-Seiberg $\Nf=1$
argument from Section 4 of Ref.~\cite{ADS} to all $\Nf < N-1$.
The relevant matter degrees of freedom are the $2\Nf^2$ classically
massless real scalars and their superpartners.
They can
be packaged into $\Nf^2$ chiral superfield (complex) degrees of freedom.
In fact, there are precisely $\Nf^2$ gauge-invariant composite chiral 
meson superfields 
$
M^r_s (x,\theta) =\Qtil^{ru}(x,\theta)  Q_{us}(x,\theta)$.
As the scalar components of the superfields $M$
have the mass-dimension two
we prefer to use an equivalent parametrisation of matter degrees
of freedom in terms of mass-dimension one chiral superfields
\begin{equation}
\Phi^r_s \ = \ {v \over \sqrt{2}} \ \log {\Qtil^r Q_s \over v^2}
\ .
\elabel{nphi}\end{equation}
The logarithm is used for the later convenience and
the mass-dimension one in \eqref{nphi} is achieved via explicit dependence 
on the classical VEV $v$ which will be necessary in the perturbative 
decoupling treatment, but will disappear from the final results.
In fact, all the non-renormalizable interactions induced by integrating
out the heavy fields will be in terms of higher dimensional operators
of the light fields divided by the powers of $v$. A perturbative 
(or semi-classical) treatment
is applicable as long as $v$ is large compared to $\Lambda_{\Nf,N}$, 
which will be assumed. There is a unique operator of dimension five
which couples the two sectors,
\begin{equation}
\Gamma \ = \ {\sqrt{2} \over 32 \pi^2} {1 \over v}
\int d^2 \theta (\Tr_\Nf \Phi) W^{a\alpha} W^a_\alpha \ + \ {\rm h.c.}
\ . 
\elabel{dimfive}\end{equation}
The the normalization factor ${\sqrt{2} \over 32 \pi^2}$ 
can be determined either from the
corresponding 1-loop perturbative supergraphs, or from requiring 
the $U(1)_{\rm R}$ invariance of the effective lagrangian
\begin{equation}\begin{split}
\Leff^{(1)} \ = \ \sum_{a=1}^{(N-\Nf)^2-1} 
\int d^2\theta \ \Big({1 \over 4 g^2 (\mu)} W^{a\alpha} W^a_\alpha \ &+ \ 
 {\sqrt{2} \over 32 \pi^2} {1 \over v}
(\Tr_\Nf \Phi) W^{a\alpha} W^a_\alpha\Big) \ + \ {\rm h.c.} \\
&+ \ 
\int d^2\theta d^2\thetabar \  
\Phi^\dagger \Phi
 \ . 
\elabel{lstone}\end{split}\end{equation}
Under  $U(1)_{\rm R}$ the pure gauge term is anomalous
\begin{equation}
\delta \ {1 \over 4 g^2}\int d^2\theta \  W^2 \ = \ 
-{(N-\Nf)\over 16 \pi^2} F^{mn} {}^*F_{mn}
\ ,
\elabel{anom}\end{equation}
and the $U(1)_{\rm R}$ transformation of $\Phi$, read from \eqref{rsym},
\begin{equation}
\delta \ \Phi^r_s \ = \ iv\sqrt{2}{(N-\Nf)\over \Nf}
\elabel{phrs}\end{equation}
leaves \eqref{lstone} invariant as required.

There are of course even higher-dimensional operators coupling the
two sectors, but they are suppressed by higher powers of $v$
and will not contribute at the relevant order.
Furthermore, perturbative non-renormalization theorems will
prevent the generation of the superpotential made solely of the
$\Phi$ fields as no such superpotential existed classically. 
Finally, note that Eq.~\eqref{lstone} should be thought of as the Wilson
effective lagrangian with $\mu$ in $g^2(\mu)$ on the right hand side
of \eqref{lstone} being the Wilson scale. In deriving $\Leff^{(1)}$
we have integrated out all the degrees of freedom with masses and virtualities
greater than $\mu$, where $\mu \le v$.

{\bf $\Nf < N-1$: Step Two}

The gauge sector together with the gauge--meson interactions sector
of the effective lagrangian $\Leff^{(1)}$ can be represented as
\begin{equation}
\Leff^{\rm gauge} \ = \ {1 \over 8 \pi} \ {\rm Im} \ 
\sum_{a=1}^{(N-\Nf)^2-1} 
\int d^2\theta \ T^{(\mu)} \ W^{a\alpha} W^a_\alpha 
\elabel{lgss}\end{equation}
where\footnote{$\Leff^{\rm gauge}$ is defined as the first line on the
right hand side of Eq.~\eqref{lstone}, i.e. $\Leff^{(1)}$ minus the
pure matter sector.} we have introduced a chiral superfield
\begin{equation} 
T^{(\mu)} \ = \ {4\pi i \over g^2(\mu)} \ + \ {1 \over 2 \pi} 
{i\sqrt{2} \ \Tr_\Nf \Phi \over v}
\ .
\elabel{tsup}\end{equation}
We now see from equations \eqref{schem} and \eqref{lgss} 
that the problem of deriving the ADS superpotential
in the theory with $N$ colours and $\Nf$ flavours is reduced
to integrating out the vector supermultiplet $W_{\st SU(N-\Nf)}$ in the
pure $SU(N-\Nf)$ supersymmetric Yang-Mills theory coupled to the
background superfield $T^{(\mu)}$
\begin{equation}
\int\D W_{\st SU(N-\Nf)}\,
e^{i\int d^4x\,\Leff^{\rm gauge}(W_{\st SU(N-\Nf)},T^{(\mu)})}
 \ = \ 
e^{ -i\left(\int d^4x\,d^2\theta \ \Weff(T^{(\mu)}) + {\rm h.c.}\right)}
\ .
\elabel{sctwo}\end{equation}
Applying the analysis of section II we conclude that
the functional integral over the gauge supermultiplet receives
contributions from $N-\Nf$ varieties of monopoles which arise
in the $SU(N-\Nf)$ supersymmetric Yang-Mills theory partially compactified
on ${\Bbb R}^3 \times S^1$. The semi-classical evaluation of the integral
in the monopole background is exact and in the decompactification limit
${\Bbb R}^3 \times S^1 \to {\Bbb R}^4$ gives, in analogy with Eq.~\eqref{sssfin}
\begin{equation}
\Weff(T^{(\mu)}) \ = \  (N-\Nf) \ \mu^3 \ \exp\Big[{2\pi i \over N-\Nf}
T^{(\mu)} \Big]
\ .
\elabel{rone}\end{equation}

In the rest of this section we want to demonstrate that the
superpotential $\Weff(T^{(\mu)})$, given by
Eq.~\eqref{rone} 
is in fact equal to the ADS superpotential $\Wads$ 
\eqref{finfin}. Indeed, substitute
the definition of $T^{(\mu)}$, Eq.~\eqref{tsup} into Eq.~\eqref{rone},
\begin{equation}
\Weff(T^{(\mu)}) \ = \  (N-\Nf) \ \mu^3 \ e^{-{8\pi^2\over g^2(\mu)} \ 
{1 \over N-\Nf}} \ \Big({v^{2\Nf} \over \det_{\Nf}\big(\Qtil^{r} Q_s\big)}
\Big)^{1 \over N-\Nf}
\ .
\elabel{rtwo}\end{equation}
Now note that the $\mu$-dependent terms combine to the RG
invariant definition of the dynamical scale of the
supersymmetric pure $SU(N-\Nf)$ gauge theory
\begin{equation}
\mu^3 \ e^{-{8\pi^2\over g^2(\mu)} \ 
{1 \over N-\Nf}} \ = \ \Lambda_{0,N-\Nf}^3
\ .
\elabel{rthr}\end{equation}
Furthermore, the decoupling relation,
$\Lambda_{\Nf,N}^{3N-\Nf}= v^{2\Nf} \Lambda_{0,N-\Nf}^{3(N-\Nf)}$
enables us to rewrite the final result in terms of the dynamical
scale $\Lambda_{\Nf,N}$ of the original SQCD:
\begin{equation}
\Weff  = \
(N-\Nf)\,\left({\Lambda^{3N-\Nf}_{\Nf,N}\over
\det_{\Nf}\big(\Qtil^{r} Q_s\big)}\right)^{1\over N-\Nf} \ = \ \Wads^{\Nf,N}
\ .
\elabel{finads}\end{equation}
Q.E.D.

\medskip

\centerline{$\sst**********************$}

\medskip

We thank Diego Bellisai and Tim Hollowood 
 for valuable discussions.
NMD acknowledges a PPARC studentship.

\end{document}